\documentclass[conference]{IEEEtran}

\usepackage[utf8]{inputenc}
\usepackage[T1]{fontenc}

\usepackage{booktabs}
\usepackage{tabularx}
\usepackage{paralist}
\usepackage{framed}
\usepackage{tikz}

\usepackage{svg}
\usepackage{url}

\usepackage{cite}
\usepackage{graphicx}

\newcommand\copyrighttext{%
  \footnotesize \textcopyright~2025 IEEE. Personal use of this material is permitted. Permission from IEEE must be obtained for all other uses, in any current or future media, including reprinting/republishing this material for advertising or promotional purposes, creating new collective works, for resale or redistribution to servers or lists, or reuse of any copyrighted component of this work in other works.}
\newcommand\copyrightnotice{%
\begin{tikzpicture}[remember picture,overlay]
\node[anchor=south,yshift=10pt] at (current page.south) {\fbox{\parbox{\dimexpr\textwidth-\fboxsep-\fboxrule\relax}{\copyrighttext}}};
\end{tikzpicture}%
}

\begin{document}




\title{From Requirements to Code: Understanding Developer Practices in LLM-Assisted Software Engineering}


\author{
    \IEEEauthorblockN{Jonathan Ullrich, Matthias Koch}
    \IEEEauthorblockA{Fraunhofer IESE \\ 
    Kaiserslautern, Germany \\ 
    \{jonathan.ullrich, matthias.koch\}@iese.fraunhofer.de}
    
    \and
    
    \IEEEauthorblockN{Andreas Vogelsang}
    \IEEEauthorblockA{paluno – The Ruhr Institute for Software Technology\\
    University of Duisburg-Essen \\ 
    Essen, Germany \\ 
    andreas.vogelsang@uni-due.de}
}

\maketitle
\copyrightnotice

\begin{abstract}

With the advent of generative LLMs and their advanced code generation capabilities, some people already envision the end of traditional software engineering, as LLMs may be able to produce high-quality code based solely on the requirements a domain expert feeds into the system. 
The feasibility of this vision can be assessed by understanding how developers currently incorporate requirements when using LLMs for code generation---a topic that remains largely unexplored. 
We interviewed 18 practitioners from 14 companies to understand how they (re)use information from requirements and other design artifacts to feed LLMs when generating code. 
Based on our findings, we propose a theory that explains the processes developers employ and the artifacts they rely on. Our theory suggests that requirements, as typically documented, are too abstract for direct input into LLMs. Instead, they must first be manually decomposed into programming tasks, which are then enriched with design decisions and architectural constraints before being used in prompts.
Our study highlights that fundamental RE work is still necessary when LLMs are used to generate code. Our theory is important for contextualizing scientific approaches to automating requirements-centric SE tasks.
\end{abstract}

\begin{IEEEkeywords}
requirements, code generation, interview study, LLM, GenAI
\end{IEEEkeywords}


\section{Introduction}

Implementation is the phase of the software engineering (SE) process in which a software design is translated into executable code~\cite{sommerville2015software}. All preceding SE activities culminate in the implementation phase. Previously elicited user and system requirements are documented and modeled in a design to be used for implementation. Requirements and design represent an abstraction of the system to be realized. They can be explicitly documented or exist implicitly in the developer's head~\cite{sommerville2015software}. Although requirements usually reside in the problem space, i.e., describing \textit{what} needs to be built, the design constitutes a bridge between the requirements and the implementation and defines \textit{how} to build the system. Nevertheless, the design does not prescribe the exact steps to realize the system in code. To do so, developers must translate a descriptive view of the system into requirements and design into actionable implementation steps. In this process, developers must consider and reflect on the requirements.

Large language models (LLMs) learn powerful representations of natural language. Since code resembles natural languages in many ways~\cite{hindle2016naturalness}, researchers applied LLMs to learn representations of code~\cite{sun2024survey}. Besides general-purpose models, such as GPT-4 or Llama, LLMs more extensively trained on code, such as Codex~\cite{chen2021evaluating}, CodeLlama~\cite{roziere2023code}, and DeepSeek-Coder~\cite{guo2024deepseek}, are proficient at many code-related tasks, such as clone and defect detection, code repair, code completion, and code summarization as evaluated on many benchmarks~\cite{wan2024deep, sun2024survey}. Tools such as GitHub Copilot offer IDE integrations for these models to further increase the ease of use by providing the model with context from the code repository.

Adopting LLMs for SE depends on the compatibility of these tools with the SE process~\cite{russo2024navigating}. Good SE processes have requirements and design phases before the implementation phase~\cite{sommerville2015software}. Much of the research, however, has not considered the role of requirements and design in LLM-assisted implementation. Indeed, requirements engineering and software design are highly underrepresented in studies~\cite{hou2024large}. On the contrary, the studies that focus on requirements as a starting point for code generation~\cite{liu2020deep, mu2024clarifygpt} do not consider actual real-world requirements (e.g., functional requirements or user stories) but rather programming tasks from coding challenges.

To address the gap in the literature regarding the use of requirements in LLM-assisted implementation in practice, we set the following research objective:

\textbf{How do practitioners incorporate requirements and design information in LLM-assisted implementation?}

In particular, we aim to find out whether practitioners use requirements as part of the prompt in the interaction with code models. Further, as requirements and implementation are connected through software design, we investigate which information on the design is relevant to be included in the prompt. 
Doing so, we contribute to a better understanding of the role of requirements engineering and design in LLM-assisted SE.

Due to the exploratory nature of this study, we employed qualitative analysis based on interviews. We conducted interviews with 18 practitioners from 14 small to large companies from 12 domains to find out how requirements and design information are reflected in the interaction with LLMs.

From the interviews, we derive a theory that describes how practitioners get \textit{from requirements to code} when using LLMs. 
In this paper, we make the following contributions:
\begin{itemize}
    \item We present a theory that describes \textbf{processes and interaction patterns} developers currently follow in practice when generating code based on requirements.
    \item Our theory also describes the \textbf{content} developers incorporate in the prompts to generate useful code, i.e., code that can be integrated in an existing code base.
    \item We provide our interview guide along with the codebook in a \textbf{replication package}\footnote{\url{https://zenodo.org/records/15005613}}.
\end{itemize}

The remainder of this paper is structured as follows: In Section~\ref{related_work}, we discuss work related to neural code intelligence and interaction with code models. We highlight the gap in the literature that this work aims to fill. In Section~\ref{study_design}, we describe our research method, the subjects of our study, and the data analysis process. In the following Section~\ref{findings}, we present the findings of our study and show the resulting process and content model. We discuss the relevance and impact of our findings in Section~\ref{discussion}. We conclude this paper in Section~\ref{conclusion_future_work} and highlight future work.

\section{Related Work} \label{related_work}

\textbf{Neural Code Intelligence}. Neural code intelligence~\cite{rabin2021understanding, sun2024survey, wan2024deep} refers to the application of deep learning techniques to train models for understanding and generating code. A recent survey~\cite{sun2024survey} identifies three evolutionary stages of neural code intelligence: embeddings, pretrained models, and LLMs. While embeddings~\cite{alon2019code2vec} and pretrained models~\cite{feng2020codebert, wang2021codet5} require task-specific training and fine-tuning, LLMs can be adapted to various tasks through prompting~\cite{brown2020language}. Several LLMs have been specifically trained in code, including Codex~\cite{chen2021evaluating}, PolyCoder~\cite{xu2022systematic}, CodeGen~\cite{nijkamp2022codegen}, StarCoder~\cite{li2023starcoder}, CodeLlama~\cite{roziere2023code}, and DeepSeek-Coder~\cite{guo2024deepseek}. These models show state-of-the-art performance in tasks such as clone detection, defect detection, code repair, summarization, and generation~\cite{wan2024deep, sun2024survey, wong2023natural}. In the following, we use the term \textit{code models} to broadly refer to neural code intelligence models leveraging LLMs.

\textbf{Generating Code from Natural Language Input}. Beyond code understanding tasks, code models are predominantly used for code generation. Natural Language to Code (NL2Code) involves translating natural language descriptions into executable code. Code models have been evaluated on NL2Code tasks using various benchmarks~\cite{wan2024deep, sun2024survey}, ranging from fundamental programming challenges~\cite{austin2021program} to more complex SE tasks~\cite{jimenez2023swe}. These evaluations primarily assess the quality of generated code based on diverse natural language inputs. Niu~et~al.~\cite{niu2023empirical} compare code models across 13 SE tasks, using natural language descriptions from the Concode dataset~\cite{iyer2018mapping} and assessing their generated Java code. In Concode, natural language descriptions are embedded in Java classes and resemble method comments (e.g., ``adds a scalar to this vector in place'' or ``increment this vector''~\cite{iyer2018mapping}). Similarly, Wang~et~al.~\cite{wang2023natural} examine NL2Code at the method level by pairing code snippets from the CodeSearchNet dataset~\cite{husain2019codesearchnet} with corresponding docstrings (e.g., ``extracts video ID from URL''~\cite{husain2019codesearchnet}).

\textbf{Requirements as Natural Language Input}. Although the above-mentioned studies focus on NL2Code, they do not consider the specifics of SE processes in which code generation occurs. In SE, requirements are the primary means of describing a system in natural language~\cite{sommerville2015software}. For code models to be effectively integrated into SE workflows, they must align with established processes such as RE~\cite{russo2024navigating,Vogelsang2024}. Liu~et~al.~\cite{liu2020deep} introduced the ``Requirements Text-based Code Generation'' (ReCa) dataset to evaluate code generation based on requirements. However, their approach treats any textual program description as a requirement, disregarding the standard definition of requirements as ``a condition or capability needed by a user to solve a problem or achieve an objective''~\cite{ieee1990standard}. The ReCa dataset, sourced from programming contest platforms, contains descriptions of programming tasks rather than genuine requirements (e.g., ``You are given an array consisting of n integers. Your task is to find the maximum length of an increasing sub-array of the given array [...] Input: [...] Output: [...]''~\cite{liu2020deep}). Mu~et~al.~\cite{mu2024clarifygpt} sought to improve LLM-based code generation through requirements clarification but also treated programming task descriptions as requirements (e.g., ``write a function to sort a list of elements''~\cite{mu2024clarifygpt}). Thus, existing NL2Code evaluations have not considered requirements as defined in SE and RE communities. 


\textbf{Developer Practices with Code Models}. Given that natural language and requirements can be ambiguous and often require clarification~\cite{Vogelsang2025}, NL2Code (or Requirements-to-Code, Req2Code) entails an interactive process with code models. Research has increasingly focused on integrating code models into user-friendly coding assistants. GitHub Copilot, a widely used in-IDE code completion tool, has been evaluated for its impact on developer productivity~\cite{ziegler2022productivity} and usability~\cite{vaithilingam2022expectation, barke2023grounded}. Vaithilingam~et~al.~\cite{vaithilingam2022expectation} assess whether GitHub Copilot helps programmers complete tasks, finding that while it does not reduce task completion time, developers appreciate it as a useful starting point. Barke~et~al.~\cite{barke2023grounded} analyze user interactions with GitHub Copilot and identify two primary engagement patterns: using it to \textit{accelerate} workflows and to \textit{explore} possible solutions. Jiang~et~al.~\cite{jiang2022discovering} conducted a user study in which participants used GenLine, a tool similar to GitHub Copilot, to solve programming tasks. Both Barke~et~al. and Jiang~et~al. show that users refine their natural language queries to optimize model responses, effectively developing a ``syntax'' for interacting with the models~\cite{jiang2022discovering, barke2023grounded}.

\textbf{Research Gap}. The aforementioned studies evaluate NL2Code and coding assistants primarily for solving programming tasks. To the best of our knowledge, no research has explored the role of requirements and design artifacts in LLM-assisted implementation in industrial practice. Aligning code models with prior SE phases is essential for integrating AI into software development processes effectively~\cite{russo2024navigating}. To address this gap, this paper aims to provide a comprehensive study of how practitioners incorporate \textit{real-world} requirements and design decisions when interacting with code models.

\section{Study Design} \label{study_design}

\begin{table*}
    \centering
    \caption{Interview Participants}
    \label{tab:interview_participants}
    \begin{tabular}{@{}llllll@{}}
        \toprule
        \textbf{Participant} & \textbf{Role} & \textbf{Company Size} & \textbf{Application Domain} & \textbf{Requirements} & \textbf{Used Interaction} \\ 
        \midrule
        P01 & Developer & Large & Service Management & User Stories & Chat \\ 
        P02 & Developer & Small & Construction Planning & Issues & IDE-Integration \\ 
        P03 & Developer & Small & Construction Planning & Issues & IDE-Integration \\ 
        P04 & Software Architect & Small & Smart City & Stand. Spec. + User Stories & Chat, IDE-Integration \\ 
        P05 & Product Owner & Large & Tax and Legal & User Stories & Chat, IDE-Integration \\ 
        P06 & Team Lead & Large & Enterprise Resource Planning & User Stories & Chat \\ 
        P07 & Team Lead & Large & Enterprise Resource Planning & User Stories & Chat \\ 
        P08 & DevOps Engineer & Small & Automotive & User Stories & Chat \\ 
        P09 & Data Scientist & Small & Smart City & User Stories & Chat \\ 
        P10 & Developer & Medium & Health & Stand. Spec. + User Stories & Chat \\ 
        P11 & Requirements Engineer & Medium & Aerospace & Stand. Spec. + User Stories & Chat, IDE-Integration \\ 
        P12 & Developer & Medium & Service Management & User Stories & Chat \\ 
        P13 & Developer & Medium & E-Commerce & User Stories & Chat, IDE-Integration \\ 
        P14 & Team Lead & Medium & E-Commerce & User Stories & Chat, IDE-Integration \\ 
        P15 & Team Lead & Small & Virtual Reality & User Stories & Chat, IDE-Integration \\ 
        P16 & Team Lead & Small & Virtual Reality & User Stories & Chat, IDE-Integration \\ 
        P17 & Developer & Large & Automotive & Stand. Spec. + User Stories & Chat \\ 
        P18 & Developer & Medium & Banking & Issues & IDE-Integration \\ 
        \bottomrule
    \end{tabular}
\end{table*}

\subsection{Research Method}

The idea for this paper started as a broad research interest: \textit{With increased use of code models in software engineering, what is the role of requirements? Are requirements, as natural language descriptions of a system, used as input for code models in practice?} Due to the nascent topic, we set a broad research focus instead of limiting our scope to predefined research questions. We employ an interview study~\cite{rubin2011qualitative} to qualitatively explore this research focus. Iteratively, we conducted interviews, analyzed the data, learned from the informants' statements, and applied the insights in subsequent interviews. In this process, we refined our research focus to include not only the role of requirements but also design information in the context of LLM-assisted software engineering. In this process, we narrowed our focus to the following research objective: \textit{How do practitioners incorporate requirements and design information in LLM-assisted implementation?}

\subsection{Interviews and Study Participants} \label{participants}

In our study, we conducted interviews with 18 practitioners from 14 companies across 12 different domains between November 2024 and February 2025. We applied convenience sampling using our network to acquire interview partners. Many of the participants in this study stated that their company is pursuing internal programs to evaluate the usage of code models in their software engineering processes. The interview partners were selected based on their experience with code models in the implementation phase. During the selection of interview partners, a diverse set of roles and domains was included. As the focal point of this study lies in using code models to generate code that meets requirements, developers were primarily chosen as interview partners, as they are the ones primarily realizing the requirements in code. Other roles, however, have also been considered to gather a broad view of the topic. In larger companies, we were able to gather a holistic view of LLM-assisted software engineering by talking to team leads and a product owner responsible for the topic (P05, P06, P07). We tried to balance breadth and depth, exploring insights from many different companies in different domains, but also collecting multiple impressions from the same company if doing so deemed to provide a deeper understanding of the usage of code models at a company (P02 and P03, P06 and P07, P13 and P14, P15 and P16 were from the same companies). We stopped acquiring new interviewees once our theory was validated in subsequent interviews and no new themes emerged. This has been the case after we conducted the interviews scheduled with P15, P16, P17, and P18. Table~\ref{tab:interview_participants} shows the participants' roles, the size of the company they work for (following the European definition of SMEs and larger firms), the application domain of the developed software, the type of requirements used by the participants (standardized specifications containing (non\mbox{-}) functional requirements, user stories, or unstructured issues) and the interface for interacting with code models (either using a chat, e.g., ChatGPT, or an IDE-integration, e.g., GitHub Copilot).

The interviews, ranging from 30 to 60 minutes, were conducted online and recorded using Microsoft Teams. The interviews were conducted by the first author following a semi-structured interview guideline. In the first part of the interview, the interviewees were asked to describe their level of experience, the software engineering process at their company, and the artifacts (i.e., documented requirements or design decisions) created in the phases before using code models for implementation. The first part of the interview concluded with the participants describing their usual implementation workflow when tasked to realize a requirement without using code models. The second part of the interview focused on the use of code models in the implementation, with a dedicated focus on aligning the generated code with requirements and design decisions. Participants were asked to outline the ``trigger'' to start using code models, the content of the used prompts, and the process by which they ensure that the generated code fulfills a requirement. All interviews were transcribed. All transcripts were translated from German to English using DeepL, except for two interviews held in English.


\subsection{Data Analysis and Theory Construction} \label{analysis_and_construction}

We inductively created a theory based on interviewees' statements by exhaustive in-vivo coding. This resulted in 179 quotes as in-vivo codes. Iteratively, the in-vivo codes were clustered, creating higher-level themes. Themes that could not be confirmed by repeatedly being present in different interviews were discarded. The initial coding was done by one of the authors using the tool MaxQDA and exported to Excel. The coding of in-vivo codes and themes was reviewed by the other two authors regarding their consistency. Collectively, the authors named the themes and discussed the resulting theory. As, in software engineering, the same phenomenon can oftentimes be described from a behavioral and structural viewpoint, we also constructed our theory concerning these two viewpoints, resulting in a process model and content model of LLM-assisted implementation based on requirements and design decisions. The replication package shows the traces from in-vivo codes to themes to derived entities in the process and content model.

\subsection{Threats to Validity} \label{threats_to_validity}

We took steps to mitigate common threats to validity in qualitative studies~\cite{creswell2017research}. 

\emph{Internal validity} refers to the threat that causal relationships or explanations are falsely drawn from the data. This can occur if the researchers' assumptions or participants’ biased responses influence the interpretation of results. To mitigate this threat and to collect all relevant data, we recorded the interviews and transcribed them word by word. We annotated the transcripts with pointers to the respective positions in the recording to trace back to the original conversation. We referred to in-vivo codes to ensure that interpretations were grounded in the actual data. To minimize the possibility of misunderstandings between interviewees and the researchers, the study goal was explained to the participants before the interview. During the interviews, we asked participants to show us intermediate artifacts such as user stories, prompt templates, and prompts, if possible.

\emph{External validity} concerns the generalizability of our findings beyond the specific context of this study. While qualitative research does not aim for statistical generalization, we addressed this threat by including participants from diverse roles, industries, and levels of experience. This variety increases the likelihood that the identified themes apply across a broader range of settings. We tried to minimize sampling bias by including practitioners from 12 different domains.

\emph{Construct validity} refers to the extent to which the interview questions and coding accurately capture the phenomena under investigation. To ensure construct validity and to mitigate researcher bias in the interpretation of the interviews, the coding was cross-validated by all authors and discussed in case of disagreements. Steps taken to further improve the reliability of the data included a review of the interview guideline.  We also discussed derived themes and models to ensure consistent interpretation of constructs.

\section{Findings} \label{findings}

In the following, we present the main findings from our study. Firstly, we describe the  \textit{process} of using code models for implementation based on requirements and design information (Fig.~\ref{fig:process_view}). Secondly, we underpin the usage of code models in LLM-assisted SE by outlining the \textit{content} that is given to a code model during this process (Fig.~\ref{fig:content_model}). We highlight the empirical evidence (i.e., participants mentioning a finding) in parentheses after each finding.

\subsection{Process Model}

\begin{figure*}
    \centering
    \includegraphics[width=\textwidth]{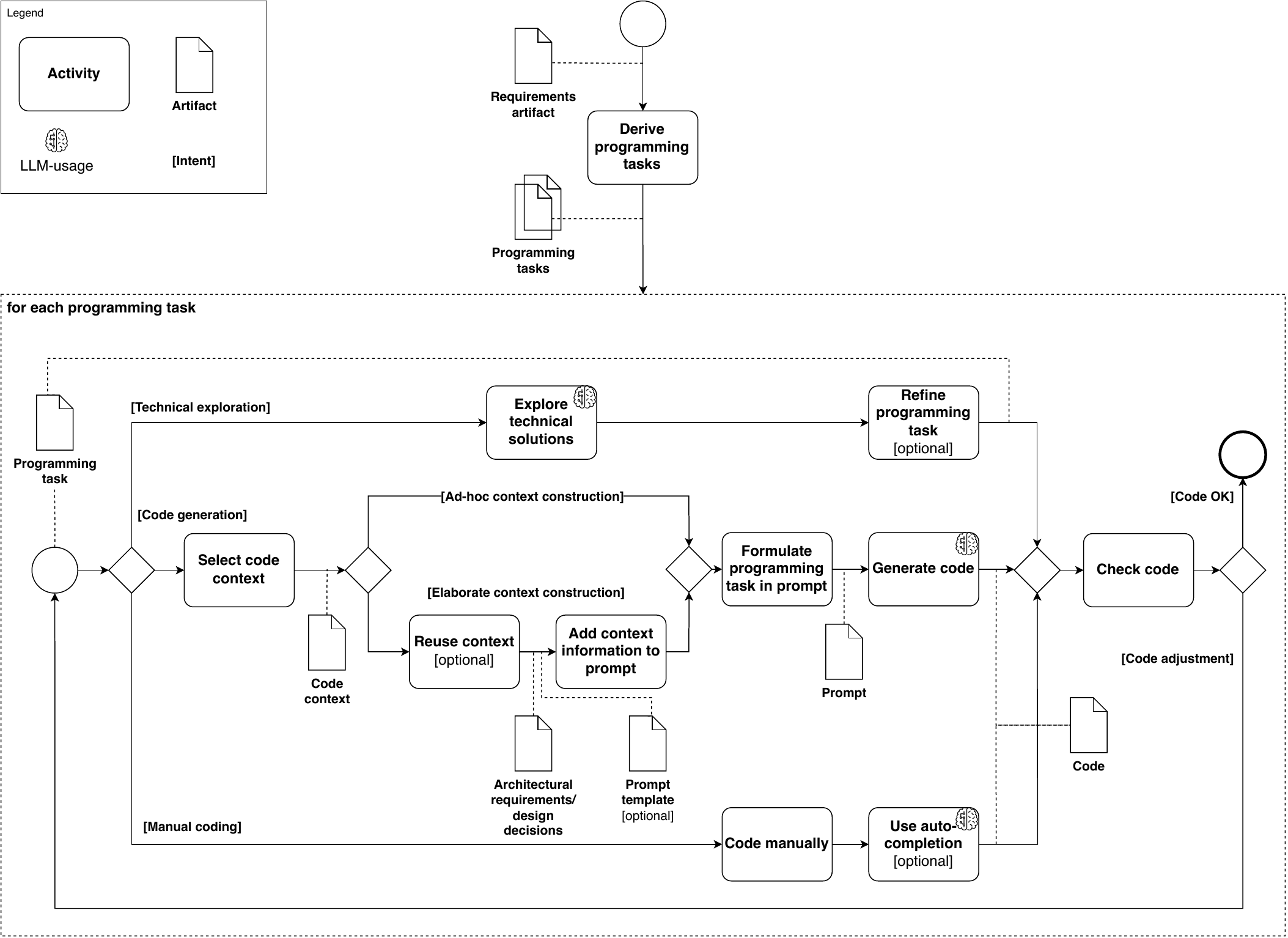}
    \caption{Process model on LLM-assisted implementation}
    \label{fig:process_view}
\end{figure*}

Practitioners engage in a process of decomposing requirements artifacts into programming tasks followed by different approaches to implement the programming tasks. We discuss Fig.~\ref{fig:process_view} in two steps: deriving programming tasks from requirements artifacts (the upper part) and the LLM-assisted implementation of programming tasks (the lower box).

\textbf{Deriving Programming Tasks From Requirements Artifacts.} 
Participants in our study stated that requirements artifacts, as traditionally documented, are too abstract to be fed into code models. Several interviewees tried using requirements artifacts as input for code models but discarded the generated code as they could not use it. P10 described this as follows: ``If I say GPT write a command interface for a Keithley [multimeter], then I have in fact taken a requirement from my catalog [...] and I've never actually had anything ready to use come out of it''. P04 shared a similar experience: ``I've found that if you try to implement entire requirements on this large level, depending on how they are formulated, then you end up with something, but you have to adapt a lot and that's not efficient. That's why there's another `breaking down' process before the prompting starts''.

Practitioners unanimously stated they first derive smaller units from the requirements artifacts. We refer to these smaller units as \textit{programming tasks}. Requirements artifacts oftentimes leave implementation details unspecified. Programming tasks include more specific details on how to realize a requirement in code. When deriving programming tasks from requirements artifacts, these details are added: ``The rough draft of the program, or how I want to structure it, I'll do that myself. I think about what I need for the remaining part and then I think about what logic to put in, which function I need to call, how to connect to the database, and so on. This structure, I do all that myself. I don't let ChatGPT do that, most of the time nothing comes out of it'' (P06). Almost all interviewees confirmed this notion of deriving programming tasks from the original requirements artifacts to use for code generation (P01, P04, P05, P06, P07, P08, P09, P10, P11, P13, P15, and P17).

P04, P05, and P06 argued that the process of refining requirements into programming tasks is not new to implementing with code models. P04 stated: ``I think this is also the classic thought process of how to implement requirements before coding assistants existed. It's not that easy to describe, because it happens subconsciously. You could now make a to-do list, and each of these to-dos would then become more or less a prompt with a feedback loop. That's probably how it would be, but it's not as if I would write all these steps down''.

P01, on the contrary, described a process of creating an explicit specification which defines the programming tasks for a user story: ``The product owner creates the [user] stories [...] then, we go there later when we edit the story and write specs on how it should work technically''. The specification supports the construction of a prompt: ``My specs say that if there is invalid input, I should handle it this way and this is one point in the spec that I'm trying to implement. Then I give the code model the existing code [and say] the behavior [from the spec] should be added'' (P01).

P04, P08, and P13 suggested that the refinement of requirements could also be done by AI: ``There would be a gap that could be closed by actually throwing the requirement to the model and then saying `first make a plan, write down these to-dos' and then you can edit them again and then implement each one individually'' (P04).

\begin{framed}
\noindent \textbf{Takeaway 1:} Practitioners do not use traditional requirements artifacts (such as user stories or functional requirements) as input for code models. Instead, they decompose and refine requirements artifacts into \emph{programming tasks} to be used as input for LLM-assisted implementation.
\end{framed}

\textbf{Implementation of Programming Tasks.} 
Our study reveals three different \textit{modes} by which practitioners use code models to implement programming tasks: \textit{technical exploration}, \textit{code generation}, and \textit{manual coding} (lower box in Fig.~\ref{fig:process_view}). Practitioners select and combine these modes depending on the familiarity of the developer with the programming task: ``There is the side of implementation itself with things that you are already familiar with. And there's the exploratory side, where you don't actually know what exactly you're supposed to do, and then I'll ask ChatGPT'' (P10). Apart from technical know-how, experience using code models also determines the chosen mode, as illustrated by a statement made by P02: ``I don't yet feel like I'm trying to do everything via code models. I still feel like a developer trying to do a lot myself and not thinking that code models could actually make some of my work easier. I'd have to practice that a bit more, I'd say''.

The \textit{technical exploration} mode is followed if a programming task is not clear enough to the developer or if the developer is uncertain about implementation details and context.
LLMs support \textit{technical exploration} by outlining solution approaches for a particular programming task. Our study participants use LLMs as an ``intelligent search engine'' (P15) or ``Google replacement that gives you a quicker and more precise answer'' (P12) during implementation. Based on the gained knowledge about possible solutions, the participants may further refine the programming task. Refinement of the programming task aims at improving the generated code: ``The further down you go, the better the result ChatGPT gives you, i.e. the more precise you describe it, and the smaller the task is, the better it actually gets a suggestion that you can use as a guide and that is really good'' (P07).

The \textit{code generation} mode is followed when code models shall generate the majority of code for a certain programming task. It is initiated by selecting the code context relevant for code generation, i.e., selecting a section of existing code that needs to be considered when generating new code. Without the code context ``there's too much knowledge on the outside that [the code model] needs'', as P06 stated, ``the simplest thing is to imagine which functions already exist where you don't have to develop them yourself and the model simply doesn't know that'' (P06). Based on this initial context for the code model, practitioners pursue different steps depending on the task and the effort they are willing to invest in constructing more specific context. For simple tasks like generating `boilerplate code` (P05), practitioners skip further context construction steps and craft the prompt based on the code context and the programming task (\textit{ad-hoc context construction}, Fig.~\ref{fig:process_view}). P01 described her workflow in this regard as follows: ``I often copy code that I already have, but now I would like [the model] to add this feature. So I already have working code, and now I would like that, when invalid input comes, it throws an error.''

Contrary to this, there are cases where more context has to be given to the model to guide the output to generate useful code (\textit{elaborate context construction}, Fig.~\ref{fig:process_view}). In these cases, information regarding design decisions and architectural constraints is added to the prompt. P05 exemplified this: ``We have certain infrastructure requirements in our company that have to be met in a certain way, that of course the co-pilot or no model knows what specifications we have for how things are to be built''. This context information has to be given in the prompt: ``The context is super important. Sometimes you have to say [...] I have an app that is based on .NET or Java, and this is a web application, and this is the controller, and the following error happens [...] anything that's not so obvious should be given, which could have an impact on the answer'' (P11).

During the interviews, participants frequently referred to chat histories and prompt templates when describing their interaction with code models. P01 stated to fall back to previous chat histories when encountering a similar problem and wants to continue from there, reusing the previous context: ``When I come across the same problem again, [...] I go back to the same prompt again and because I want it to remember all the stuff we've already discussed and now I want to add something''. P06, P07, P08, and P11 use prompt templates for different implementation tasks. P13 and P17 set up `Agents' or `Custom GPTs' in their tooling, essentially pre-filling the context with information for repetitive tasks. Reusing chat histories, using prompt templates, and pre-filling context information reduce the effort to make the generated code useful in a certain context. Other interviewees confirmed this notion of different amounts of effort put into the formulation of the prompt (P06, P08, P09, P10, P13, P17).

All participants denied documenting the prompts for later reference. P11 replied: ``I'm not doing that at the moment [...] It certainly wouldn't be bad, although I don't know whether you really look into it or not. So I think what helps more are prompt templates, i.e., saying that I have certain standard prompts and then maybe I can just fill in certain placeholders, like a form''.

The \textit{manual coding} mode is followed when developers prefer to start coding manually or the generated code needs to be adjusted. The majority of interviewees stated that the generated code always needs some manual adjustment. Some even said they prefer to start coding manually and then integrate generated code suggestions as they go (\textit{Manual coding}, Fig.~\ref{fig:process_view}): ``I either keep writing because the suggestion doesn't make sense or I just use the auto-complete suggestions and sometimes adjust them a bit'' (P18). Regardless of the selected mode, the code must be inspected to check whether it realizes the programming task or the code needs to be adjusted. Practitioners iterate through the depicted process multiple times until they have implemented all programming tasks to realize a requirement.

\begin{framed}
\noindent \textbf{Takeaway 2:} Practitioners implement programming tasks by a combination of manual coding and code generation. When generating code, they try to reduce the effort of adding relevant context information to prompts by using prompt templates, reusing contexts from chat histories, or pre-filling context information.
\end{framed}

\textbf{Interaction Patterns.} 
Practitioners reported different forms of iterating through the depicted process. The most commonly reported practice is using code models as a 'pair programmer' in a pattern we call \textit{incremental code generation}. Practitioners follow all paths in the process shown in Fig.~\ref{fig:process_view}. They interleave technical exploration, manual coding, and code generation. In this process, they incrementally construct context that they reuse when needed. This pattern is prevalent in reports of practitioners using a chat interface as it allows for iterative interaction and the reuse of the chat history (i.e., context information). Practitioners implement a programming task by manually adjusting the generated code or by reformulating the prompt.

In contrast to the aforementioned pattern, interviewees P02, P03, P05, and P18 described a pattern that can be summarized as \textit{manual coding with intelligent auto-completion}. Here, code models have little impact on the practitioners' workflow itself: Interviewees reported writing code as they normally would without code models but accept generated auto-completions and code suggestions, if applicable. Practitioners have mentioned this pattern mainly when using IDE-integrated code models. They spend most of the time in the 'manual coding' mode in Fig.~\ref{fig:process_view}. As the code model's generated auto-completions are inserted where developers normally insert code, developers can assess the code with little effort. In this interaction pattern, developers rely less on the ability of code models to combine multiple implementation steps. Instead, they interact with code models at a lower level of abstraction when accepting code suggestions to implement a programming task.

Lastly, interviewees P04, P08, P13, and P14  stated that they use code generation whenever possible. We call this pattern \textit{extensive code generation}, where practitioners offload a larger portion of the programming task to a code model. Practitioners who employ this form of interaction spend most of the time (re)formulating the prompt and checking whether the generated code fulfills the desired functionality: ``I would say I don't really write code myself anymore. I mostly iterate with AI'' (P13).

Generally, our study indicates a trade-off between the effort invested in formulating prompts and the efficiency gained from generating code. P12 argued that: ``If I choose every prompt super specifically, then I'll get it right, but then I don't need the AI either [...] If I had to think about every little thing myself, if I had to write it [in the prompt] myself, then I wouldn't need the AI to help me in any way''. 

Respondents using \textit{manual coding with intelligent auto-completion} referred to lower level forms of interaction (closer to the code), whereas interaction with code models in \textit{incremental} or \textit{extensive code generation} might be more abstract, leveraging the capability of code models to combine multiple implementation steps in one output. From this finding, we derive the following takeaway:

\begin{framed}
\noindent \textbf{Takeaway 3:} The level of abstraction of a programming task varies depending on the interaction pattern. Practitioners break down requirements accordingly into programming tasks to accommodate a certain interaction pattern.
\end{framed}

\subsection{Content Model} \label{content_view}

\begin{figure}
    \centering
    \includegraphics[width=\columnwidth]{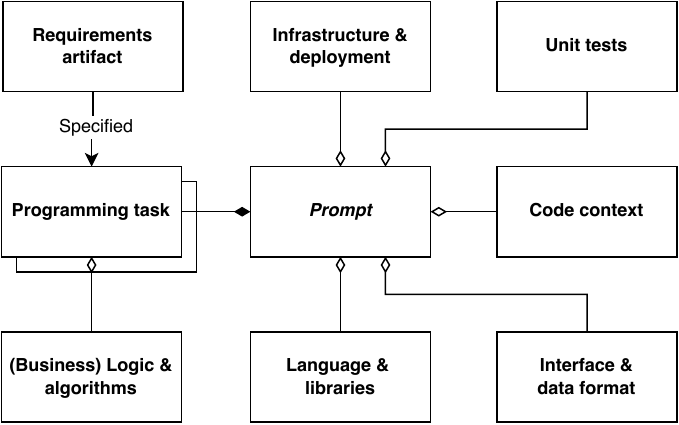}
    \caption{Content model on LLM-assisted implementation}
    \label{fig:content_model}
\end{figure}

In the content model (Fig.~\ref{fig:content_model}), we focus on the information from requirements and design decisions that is incorporated into the interaction with code models at different points in the process. One participant stated the motivation to add specific information about the constraints as follows: ``You have to expand the context accordingly. All architectural boundary conditions really have to be documented in some way. And that goes right down to the skill sets of the individual developers, because they also influence architectural decisions [...] I think that would be the first step, i.e., to really expand the functional requirements with all the architectural conditions and then to expand them in the context of what is already there, so that it remains consistent with the previous code base'' (P04). In this statement, P04 highlighted how (functional) requirements act as a starting point for LLM-assisted implementation but are only useful for code generation if enriched with information on specific design decisions. Other study participants also emphasized the importance of specifying architecturally significant requirements~\cite{chen2013Characterizing} and design decisions in the interaction with code models. Practitioners add information to the prompt context to ensure the generated code can be used, i.e., the code can be integrated into an existing code base. P06 stated this as follows: ``All these boundary conditions that we have [...] flow into [the prompt]. For example, if it is already clear that we want to use a certain programming language or a framework, then that is included [...] because otherwise I might end up with a result that works on its own, but isn't at all useful in our context''. P05 added to this: ``When I say cloud-native, that's a standing term for us, where we say we have a certain idea of what an application should look like, what environment it should run in, what basic systems we have, how things should fit together and how they have to be built. And these are just typical requirements that you can take into account, if you say I have a tool that generates code for me''.

Fig.~\ref{fig:content_model} shows an overview of the content that interviewees repeatedly reported to include into the prompt for \textit{code generation} (`add context information to prompt' in Fig.~\ref{fig:process_view}). From the interviews, we find that information related to \textit{infrastructure and deployment} (P04, P05, P06, P08), \textit{unit tests} (P05, P06, P07, P10, P17), \textit{interface and data format} (P03, P06, P09, P10), and \textit{language and libraries} (P01, P02, P03, P04, P05, P07, P09, P12, P13, P15) is added to the prompt to ensure the generated code can be integrated into an existing code base. \textit{Infrastructure and deployment} refers to ``infrastructure requirements [...] that have to be met in a certain way that, of course, Copilot or no model knows'' (P05). \textit{Unit tests} and information about the \textit{interface and data format} are added to the context as a way to guide the code models' output. While the used \textit{programming language and libraries} can oftentimes be inferred from the code context, study participants mentioned outdated information about libraries and missing information about domain-specific languages to limit the usefulness of the generated code. The \textit{programming task} is the core component in the prompt and may include information about the \textit{business logic and algorithms} to be used (as stated by P01, P02, P03, P09, P11, P13, and P16). 

\begin{framed}
\noindent \textbf{Takeaway 4:} To craft a good prompt around a given programming task, practitioners add specific business logic or algorithmic information along with constraints on infrastructure and deployment, languages and libraries, interfaces and data formats, unit tests, and the code context in which the generated code shall be integrated.
\end{framed}

\section{Discussion} \label{discussion}
Our study provides answers to our research objective: \textit{How do practitioners incorporate requirements and design information in LLM-assisted implementation?} The findings suggest that constraints from architecturally significant requirements~\cite{chen2013Characterizing} or design decisions are directly used by developers in the interaction with code models; more abstract user-level requirements are not suited for direct input to generate useful code. Information from traditional requirements artifacts (such as user stories or standardized specifications containing functional requirements) is only indirectly incorporated in LLM-assisted implementation in practice. Practitioners manually decompose traditional requirements artifacts into concrete programming tasks to be included in the prompt. Design decisions and architecturally significant requirements (e.g., non-functional requirements) act as constraints for the generated code and are directly incorporated in the interaction with code models (i.e., specified in the prompt). There is, however, no consistent general approach to incorporating requirements and design information in LLM-assisted implementation. Our study contributes towards classifying different modes of interacting with code models. In the following, we discuss these findings in relation to existing evidence and how these impact research and practice.

\textbf{Relation to existing evidence.} Our findings can be related to several findings from previous studies. Barke~et~al.~\cite{barke2023grounded} identified two primary engagement patterns of developers when using LLMs for code generation: using LLMs to \textit{accelerate} workflows and to \textit{explore} possible solutions. Participants in our study confirmed this notion. As shown in Figure~\ref{fig:process_view}, practitioners engage in activities to accelerate their work ('generate code' and 'use auto-completion') or to 'explore technical solutions'. Our study puts Barke~et~al.'s findings into the perspective of the overall implementation process, including requirements and design artifacts. 

Russo~\cite{russo2024navigating} emphasizes that ``the adoption of Generative AI tools hinges largely on their successful integration with existing software development workflows''~\cite{russo2024navigating}. By outlining the process of LLM-assisted implementation, as reported by our study participants, we contribute to understanding when and how LLMs are used in existing SE processes in practice. Further, Russo found ``concerns about the limitations of LLMs, such as their lack of knowledge about internal APIs and the need for human oversight, emphasize the importance of human expertise in utilizing these tools effectively''~\cite{russo2024navigating}. Our results confirm these findings in the sense that participants in our study also reported the need for searching and adding project-specific contextual information to prompts. They also unanimously reported that the resulting code must be checked and adapted in most cases. By outlining the content of a prompt in LLM-assisted implementation, we describe what information LLMs are missing that needs to be added manually.

Xu et al.~\cite{xu2022ide} study the usage of IDE-integrated assistants for NL2Code tasks before the advent of LLMs. They evaluated an IDE plugin that takes a natural language query and outputs a list of generated code snippets and retrieved Stack Overflow answers. 
They already hypothesized that integrating context information into the queries is necessary for future work. Our results support this hypothesis and strengthen it further. Not only does the local code context play a crucial role, as hypothesized by Xu~et~al., but also additional information, such as architectural requirements and design decisions, is relevant for generating useful code. 

We emphasize the role of programming tasks as technical specifications of how to implement a requirement in multiple steps in LLM-assisted implementation. A qualitative study by Liang et al.~\cite{liang2023qualitative} focuses on ``implementation design decisions'', i.e., decisions made by developers on how to implement a certain design when there are potential alternatives. Their study shows that developers constantly monitor high-level requirements and design decisions during implementation. However, they find that implementation decision-making processes are not consistent across developers. Our study outlines a process model and suggests interaction patterns used by developers. Similar to Liang et al. we also observed that practitioners do not follow a general structure to incorporate requirements-related information in their interaction with code models. Further research would need to identify conditions that influence developers to choose one or the other interaction pattern.

\textbf{Impact for research.}
We believe our results contribute to contextualizing research on automatic requirements-centric SE tasks. By ``contextualizing'', we mean evaluating approaches in light of current developer practices and the challenges they present. Our findings indicate that requirements, as documented in practice today, are rarely suitable for direct use as prompt inputs. Instead, practitioners decompose them into smaller tasks and enrich them with contextual information to generate useful code. Any approach that claims to automate a requirements-centric SE task (e.g., code generation~\cite{mu2024clarifygpt,wei2024requirements}, test case generation~\cite{Arora24}, or model generation~\cite{ferrari2024}) should explicitly discuss the representativeness of its requirements and how it addresses the specific demands for producing useful code in practice. Researchers should carefully distinguish between NL2Code and Req2Code (Requirements-to-Code) approaches.

Additionally, our findings suggest the need for further research on supporting practitioners in the manual steps outlined in our theoretical framework. One particularly interesting insight from our study is the significant role of decomposing requirements into programming tasks that are appropriately sized for LLM prompts. The key question is determining the optimal level of granularity for prompts. Research on this topic might benefit from comparing the level of granularity of the natural language input used in previous studies on NL2Code and the level of granularity of programming tasks derived from requirements in practice.

Another research direction emerging from our results involves helping practitioners construct and reuse context. LLM context, such as in chat histories, appears to be a crucial element to generating useful code. Practitioners recall similar tasks from the past, revisit the corresponding chat history, and reuse that context for new tasks. Systematically supporting the construction and reuse of prompt contexts (i.e., leveraging existing design documentation or stored chat histories for future tasks) remains an underexplored area of research.

A third research avenue is investigating the trade-off between investing effort in prompt engineering versus adapting the generated code. Participants in our study observed that the quality of generated code improves with more carefully crafted prompts. However, since the output is never perfect and always requires post-generation modifications, the effort spent on prompt engineering must be balanced against the effort required for code adaptation. One participant noted that when time is limited, they invest less effort in prompt engineering, which can lead to increased rework after code generation. Managing this trade-off remains an open question, closely related to determining what constitutes ``good enough'' requirements engineering~\cite{Fricker2019,Frattini2024,Femmer2019}.

\textbf{Impact for practice.}
Based on our findings, we conclude that the vision of fully automated software engineering---where domain experts, without necessarily possessing technical expertise, simply articulate their needs and desires to an LLM, which then produces a complete application---is still a distant reality, particularly for more complex software. Our results indicate that while LLMs are actively used in practice for code generation, the process is preceded by several manual and creative steps that require expertise in requirements and software engineering.

Our interview participants confirmed that requirements, as documented today, are often too abstract and must be broken down---an essential task in requirements engineering. Similarly, they emphasized that identifying relevant contextual information and incorporating it into a prompt is crucial for obtaining useful code. This ability, which is fundamental to requirements and software engineering, is difficult for non-technical experts to perform effectively.

Additionally, our findings reveal that prompts are currently not regarded as important engineering artifacts that require documentation, storage, traceability, or review. It would be valuable to assess the implications of this transient use of prompts. Since the quality and relevance of generated code depend on the input provided in the prompt, we hypothesize that understanding which prompt produced a given output and tracing the origins of prompt information will become increasingly relevant---especially in domains where accountability and correctness are critical.

Therefore, we conclude that requirements and software engineering activities and skills remain essential in contemporary software development projects, even when LLMs are used for code generation.

\section{Conclusion} \label{conclusion_future_work}

In this paper, we propose a theory that contextualizes requirements, artifacts, and design decisions in LLM-assisted implementation. This theory is based on semi-structured interviews with 18 practitioners from 14 companies across 12 domains. We contribute to the understanding of LLM-assisted software engineering by describing how requirements and design decisions are incorporated into the LLM-assisted implementation process. Our findings show that practitioners break down requirements into programming tasks to serve as input for code models. They also integrate information from architectural requirements and design decisions to ensure the generated code aligns with the existing code base. Additionally, we outline different interaction patterns and the composition of prompts in LLM-assisted implementation. We discuss the implications of these findings for requirements engineering research and practice.

Our study provides a foundation for future research on key activities in LLM-assisted implementation. Further studies could explore the decomposition of requirement artifacts into programming tasks in greater detail or examine how context information can be retrieved from existing documentation based on the prompt content model proposed in this study.

\section*{Acknowledgments}
We thank our interview participants for their time and commitment to contributing to our study. We also thank the reviewers for their extensive feedback and suggestions to improve this paper.

\bibliographystyle{IEEEtran}
\bibliography{references}
\end{document}